\documentstyle[12pt,a4,epsf]{article}

\def\gsim{\displaystyle\mathop{>}_{\sim}}
\def\lsim{\displaystyle\mathop{<}_{\sim}}

\newcommand{\be}{\begin{equation}}
\newcommand{\ee}{\end{equation}}
\newcommand{\bea}{\begin{eqnarray}}
\newcommand{\eea}{\end{eqnarray}}
\newcommand{\Psb}{\bar{\Psi}}
\newcommand{\psb}{\bar{\psi}}

\newcommand{\chb}{\bar{\chi}}
\newcommand{\nb}{\bar{N}}
\newcommand{\bb}{\bar{B}}
\newcommand{\pa}{\partial}
\newcommand{\lto}{\longrightarrow}
\newcommand{\tr}{{\rm Tr}}
\newcommand{\sla}{\hspace{-0.5em} /}

\begin{document}
\begin{flushright}
  TIT/HEP-379/NP
\end{flushright}

\begin{center}
{\Large \bf Decays of $\frac{\bf 1}{\bf 2}^{\bf -}$ Baryons in Chiral 
Effective Theory}

\vskip 10mm
Y. Nemoto\footnote{E-mail address: nemoto@th.phys.titech.ac.jp},
D. Jido and 
M. Oka \\ 
   {\it Department of Physics, Tokyo Institute of Technology} \\
   {\it Meguro, Tokyo 152, Japan}
 \vskip 5mm
A. Hosaka \\ 
   {\it Numazu College of Technology} \\
   {\it 3600 Ooka, Numazu 410, Japan} 
\end{center}

\baselineskip 1.5em

\begin{abstract}
\baselineskip 1.5em
  We construct an $SU(3)_L \times SU(3)_R$ symmetric chiral effective model 
which includes parity pair baryon fields.
It is assumed that the positive and negative parity baryons in the parity 
pair have opposite chiral transformation properties each other.
Using this model, strong decays of the negative parity baryons are studied
up to the order of one-loop corrections.
The results agree qualitatively with experiment in the $\pi$ and $K$ 
channels.
We examine how the model parameters are determined from the decay widths
and study their physical meanings.
Possibilities to build other models and relations to the
correlation function analysis are discussed.
\end{abstract}

PACS number(s): 11.30.Rd, 12.39.Fe, 13.30.-a

\section{Introduction} \label{sc:intro}

 Chiral symmetry is one of the most important features in the low-energy
QCD and deeply influences low-lying hadron properties.
In the meson sector, eight light pseudoscalar mesons, $\pi$, $K$ and $\eta$,
are regarded as the Nambu-Goldstone(NG) bosons associated with the spontaneous 
chiral $SU(3)_L \times SU(3)_R$ symmetry breaking(SCSB).
The symmetry also tells us that these mesons satisfy low-energy theorems
which lead to predictions such as the Goldberger--Treiman relation, the 
Gell-Mann--Oakes--Renner relation and so on.

  In raising temperature, as shown in lattice QCD calculation, the chiral 
symmetry is believed to be restored at around $T_C \sim m_\pi$.
Then the pion, which is the NG-boson before the chiral restoration, becomes
a member of an irreducible representation $(2,2)$ of $SU(2)_L \times SU(2)_R$
with a degenerate scalar partner, $\sigma$.

  As for baryons, when the chiral symmetry is restored, each baryon should have
degenerate parity partner and form a parity pair.
Mass splittings between the parity partners in the real world would emerge 
as a result of spontaneous and explicit chiral symmetry breaking.
It is, however, not known well as compared with mesons how the SCSB works
quantitatively on the properties of the baryons.

  Two of the present authors(D.J and M.O) and Kodama previously studied the 
masses of the flavor octet and singlet negative parity $\frac{1}{2}^-$ 
baryons using the QCD sum rule\cite{JO96a}.
There the same interpolating field is used for the positive and 
negative parity baryons and their masses are evaluated.
Then they study the relation between the positive and negative parity 
baryons with an emphasis on chiral symmetry.
As a result, the masses of the lowest-lying negative parity 
baryons are fairly well reproduced using the standard values of QCD 
parameters and condensates.
One of notable findings is that the chiral-odd vacuum condensates which 
break the chiral symmetry, 
e.g., $\langle \bar{q} q \rangle$ and $\langle \bar{q} g \sigma 
\cdot G q \rangle$, cause the mass splitting of the positive and negative 
parity baryons.
Furthermore it is also pointed out that as those condensates
decrease, the masses of the negative parity baryons approach
those of the positive parity baryons, and eventually
they become degenerate when the chiral symmetry is restored.
This implies that the lowest-lying negative parity baryons can be regarded
as parity partners of the positive parity baryons of ground states.

  The main purpose of the present work is to study properties of the negative 
parity baryons as members of parity pairs from SCSB point of view.
We investigate strong decays of the flavor-octet negative parity baryons 
using chiral effective models.
The $SU(2)_L \times SU(2)_R$ linear sigma model with parity pair has been
introduced by several authors so far in somewhat different 
ways\cite{DK89}-\cite{Zh92}.
We generalize the $SU(2)_L \times SU(2)_R$ model based on DeTar and 
Kunihiro(DK) to $SU(3)_L \times SU(3)_R$.
DK defines a parity pair that consists of  a usual fermion field and 
a ``mirror fermion'' field\cite{Mo87} which is a fermion field 
whose left- and right-handed transformation property is reversed.
The $SU(3)$ linear sigma model, however, cannot be a straightforward extension 
of the $SU(2)$ linear sigma model since the baryon fields in the 
$SU(3)$ symmetry do not belong to a fundamental representation but to an 
adjoint representation.
In order to avoid an ambiguity in defining a linear baryon field,
we construct a nonlinear effective Lagrangian.
In the  nonlinear theory, 
one can incorporate effects of the $SU(3)$ symmetry 
breaking systematically by using the chiral perturbation method.

  In Sec.\ref{sc:su2}, we consider how the parity pair nucleons can be 
introduced in the effective theory.
We show that if the negative parity nucleon field transforms in the same 
way as the positive parity nucleon field, 
they decouple each other, that is, the Yukawa coupling of the pions between 
positive and negative parity baryons vanishes.
Next we introduce an $SU(2)$ parity pair linear sigma model with
a mirror fermion field according to DK.
In this case the Yukawa coupling between the positive and negative parity 
nucleons becomes nonzero.
Furthermore the masses of nucleons remain finite when the chiral symmetry
is restored.
In Sec.\ref{sc:su3}, the $SU(2)$ model is extended to $SU(3)$,
and the nonlinear effective Lagrangian is constructed.
In Sec.\ref{sc:decay}, we apply the effective theory 
to strong decay processes of negative parity baryons.
Our interests are two fold: one is the relatively suppressed couplings with
the pion, and the other is the substantial strength of the couplings with
the $\eta$ meson.
Here we study what causes these characteristic features from the viewpoint of
the chiral symmetry.
Comparison with other models is briefly given in Sec.\ref{sc:discuss}.
Relations to the analysis by the correlation function analysis\cite{JO96b} 
are also discussed here.
Summary and conclusion are given in Sec.\ref{sc:concl}.

\section{$\bf SU(2)_L \times SU(2)_R$ parity pair models: A naive vs mirror
construction}  \label{sc:su2}

 In this section, we consider how baryon fields in the parity pair should be
transformed under chiral transformations using an $SU(2)_L \times SU(2)_R$
linear sigma model.
One assumes that the nucleon fields are defined as
\be
  \Psi = \left( 
  \begin{array}{c}
    \psi_+ \\ \psi_-
  \end{array}
  \right)
\ee
where $\psi_+$ and $\psi_-$ denote the fields of the positive and 
negative parity nucleons, respectively.
We assume that in the linear sigma model, $\psi_+$ and $\psi_-$ transform
linearly under chiral transformation.
There are two possibilities for the chiral transformations of $\psi_+$
and $\psi_-$.
First we consider what we call the naive construction in which
$\psi_+$ and $\psi_-$ transform in the same way under 
$SU(2)_L \times SU(2)_R$ as
\[
  \psi_{+L} \lto \psi'_{+L} = L \psi_{+L}, \ \ \ \ \  
  \psi_{+R} \lto \psi'_{+R} = R \psi_{+R}
\]
\be
  \psi_{-L} \lto \psi'_{-L} = L \psi_{-L}, \ \ \ \ \  
  \psi_{-R} \lto \psi'_{-R} = R \psi_{-R}
  \label{eq:su2tr1}
\ee
\[
  \ \ \ \ \ \  L \in SU(2)_L, \ \ \   R \in SU(2)_R
\]
where $\psi_L = \frac{1-\gamma_5}{2} \psi$ and $\psi_R = \frac{1+\gamma_5}{2}
\psi$ denote
the left- and right-handed components of the nucleon fields, respectively.
Using these fields, the chiral invariant Lagrangian of the 
$SU(2)_L \times SU(2)_R$ linear sigma model is given by
\be
  {\cal L} = \Psb i \pa \sla \Psi 
  - g_1 \Psb (\sigma + i \gamma_5 \vec{\tau} \cdot \vec{\pi}) \Psi
  + g_2 \Psb \rho_3 (\sigma + i \gamma_5 \vec{\tau} \cdot \vec{\pi}) \Psi
  - \mu_0 \Psb i \gamma_5 \rho_2 (\sigma + i \gamma_5 \vec{\tau} \cdot 
  \vec{\pi}) \Psi + {\cal L}_M
  \label{eq:su2lag1}
\ee
where ${\cal L}_M$ is a meson part of the Lagrangian and $\rho$ and $\tau$ 
are the Pauli matrices in the parity pair space and the isospin space,
respectively.
The fourth term of RHS in eq.(\ref{eq:su2lag1}) gives a coupling of 
$\psi_+$ and 
$\psi_-$ and ensures the parity of $\psi_-$ against that of $\psi_+$.
If SCSB takes place and $\sigma = \langle 0 | \sigma | 0 \rangle
+ \tilde{\sigma}$, the mass matrix is
\be
  \sigma_0 \left( 
  \begin{array}{cc}
    g_1 - g_2     &  \mu_0 \gamma_5 \\
    -\mu_0 \gamma_5 &  g_1 + g_2
  \end{array}
  \right)
\ee
with $\sigma_0 = \langle 0 | \sigma | 0 \rangle$.
This can be diagonalized by the eigenfunction
\[
  N \equiv \left(
  \begin{array}{c}  N_+ \\ N_-  \end{array}
  \right) = \frac{1}{\sqrt{2 \cosh \delta'}} \left(
  \begin{array}{cc}
   e^{\frac{\delta'}{2}}           &  e^{-\frac{\delta'}{2}} \gamma_5 \\
   e^{-\frac{\delta'}{2}} \gamma_5 &  -e^{\frac{\delta'}{2}}
  \end{array}
  \right) \Psi
\]
\be
  \bar{N} = \Psb \frac{1}{\sqrt{2 \cosh \delta'}} \left(
  \begin{array}{cc}
   e^{\frac{\delta'}{2}}            & -e^{-\frac{\delta'}{2}} \gamma_5 \\
   -e^{-\frac{\delta'}{2}} \gamma_5 & -e^{\frac{\delta'}{2}}
  \end{array}
  \right)
  \label{eq:npnm}
\ee
with $\sinh \delta' = g_1 / \mu_0$ and the eigenvalues are
\be
  m_\pm = \sigma_0 (\mp g_2 + \sqrt{g_1^2 + \mu_0^2})
\ee
for $N_+$ and $N_-$, respectively.
We note that $N_+$ and $N_-$ are parity eigenstates as ensured by the 
$\gamma_5$ matrix in eq.(\ref{eq:npnm}). 
We see that $g_2$ should be positive so that the negative parity nucleon 
is heavier than the positive parity nucleon.
Furthermore, we obtain $m_\pm = 0$, if the condensate $\sigma_0$ vanishes
as the chiral symmetry is restored.
Now the Lagrangian (\ref{eq:su2lag1}) in terms of $N$ is 
\be
  {\cal L} = \nb i \pa \sla N - \nb m N - \nb \frac{m}{\sigma_0}
  (\tilde{\sigma} + i \gamma_5 \vec{\pi} \cdot \vec{\tau}) N
  \label{eq:su2lag2}
\ee
\[
  m = \left( \begin{array}{cc} m_+ & 0 \\ 0 & m_- 
  \end{array} \right) = \sigma_0 
  \left( \begin{array}{cc} -g_2 + \sqrt{g_1^2+\mu_0^2} &
  0 \\ 0 & +g_2 + \sqrt{g_1^2+\mu_0^2} \end{array} \right)
\]
This shows that $N_+$ and $N_-$ decouple from each other, 
as the Lagrangian is nothing but the sum of the linear sigma models for
two independent fermions.
This property is unchanged when one extends the $SU(2)_L \times SU(2)_R$
linear sigma model to the $SU(3)_L \times SU(3)_R$
model and further to nonlinear realizations.
Though this Lagrangian can be a candidate for an effective model,
it gives no significant relations between $N_+$ and $N_-$,
that is, the coupling of $N_-$ to $N_+$ and a meson is completely missing
which apparently contradicts experiments.
Therefore, we choose here another model which allows these couplings.

   According to DK\cite{DK89},
let us introduce the negative parity nucleon field as a ``mirror fermion" field
which transforms in the opposite way as the positive parity nucleon field 
under $SU(2)_L \times SU(2)_R$.
The transformation rules are
\[
  \psi_{+L} \lto \psi'_{+L} = L \psi_{+L}, \ \ \ \ \  
  \psi_{+R} \lto \psi'_{+R} = R \psi_{+R}
\]
\be
  \psi_{-L} \lto \psi'_{-L} = R \psi_{-L}, \ \ \ \ \  
  \psi_{-R} \lto \psi'_{-R} = L \psi_{-R}
  \label{eq:su2tr2}
\ee
The chiral invariant Lagrangian is then
\be
  {\cal L} = \Psb i \pa \sla \Psi 
  - g_1 \Psb ( \sigma + i \vec{\pi} \cdot \vec{\tau} \rho_3 \gamma_5 ) \Psi
  + g_2 \Psb ( \rho_3 \sigma + i \vec{\pi} \cdot \vec{\tau} \gamma_5 ) \Psi
  - i m_0 \Psb \rho_2 \gamma_5 \Psi
  + {\cal L}_M
  \label{eq:su2lag3}
\ee
Now the term proportional to $m_0$ which mixes $\psi_+$ and $\psi_-$ is
characteristic.
Since it is chirally invariant without the meson fields, it gives rise to 
a nonzero nucleon mass when the chiral symmetry is restored.

  The mass matrix of eq.(\ref{eq:su2lag3}) is given by
\be
  \left( 
  \begin{array}{cc}
    (g_1 - g_2) \sigma_0     &  m_0 \gamma_5 \\
    -m_0 \gamma_5            &  (g_1 + g_2) \sigma_0
  \end{array}
  \right)
\ee
and is diagonalized by the nucleon field
\[
  N = \left( \begin{array}{c} N_+ \\ N_- \end{array} \right) = 
  \frac{1}{\sqrt{2 \cosh \delta}} \left(
  \begin{array}{cc}
   e^{\frac{\delta}{2}}           &  e^{-\frac{\delta}{2}} \gamma_5 \\
   e^{-\frac{\delta}{2}} \gamma_5 &  -e^{\frac{\delta}{2}}
  \end{array}
  \right) \Psi
\]
\be
  \bar{N} = \Psb \frac{1}{\sqrt{2 \cosh \delta}} \left(
  \begin{array}{cc}
   e^{\frac{\delta}{2}}            & -e^{-\frac{\delta}{2}} \gamma_5 \\
   -e^{-\frac{\delta}{2}} \gamma_5 & -e^{\frac{\delta}{2}}
  \end{array}
  \right)
\ee
with $\sinh \delta = g_1 \sigma_0 / m_0$.
Then $N_+$ and $N_-$ have the masses $m_+$ and $m_-$, respectively,
\be
  m_\pm = \mp g_2 \sigma_0 + \sqrt{(g_1 \sigma_0)^2 + m_0^2}
\ee
The vacuum expectation value $\sigma_0$ causes the mass splitting of $N_+$ and 
$N_-$.
If the chiral symmetry is restored, $\sigma_0 = 0$,  
the two masses coincide, $m_\pm = m_0$.

  Let us briefly look at phenomenological consequences of the present model.
One of interesting features of the present model is the axial property of the
nucleons.
The axial-vector coupling constants are given by
\be
  g_A \equiv \left( \begin{array}{cc}
    g_{AN_+ N_+} & g_{AN_+ N_-} \\
    g_{AN_+ N_-} & g_{AN_- N_-}
  \end{array} \right) = \left( \begin{array}{cc}
    \tanh \delta            & -\frac{1}{\cosh \delta} \\
    -\frac{1}{\cosh \delta} & -\tanh \delta
  \end{array} \right)
\ee
Note that the diagonal axial-vector coupling constants of the nucleon 
decrease as $\sigma_0$ decreases when the chiral symmetry is restored.
Then they satisfy the Goldberger--Treiman relations with the
pion-nucleon coupling constants.
These are at tree level given by
\be
  g_{\pi N_+ N_+} = g_{AN_+ N_+} \frac{m_+}{\sigma_0},\ \ 
  g_{\pi N_- N_-} = g_{AN_- N_-} \frac{m_-}{\sigma_0},\ \ 
  g_{\pi N_+ N_-} = g_{AN_+ N_-} \frac{m_+ - m_-}{2\sigma_0}
\ee
In \cite{DK89}, the negative parity nucleon is assigned to the lowest 
energy $\frac{1}{2}^-$ state, $N(1535)$, and
the parameters are determined from its pionic decay, i.e., the decay
width for $N(1535) \to N + \pi$ and the masses of $N$ and $N(1535)$.
Using $\sigma_0 = f_\pi = 93$MeV, $\Gamma_{N(1535)\to N \pi}\simeq 70$MeV,
$m_+ = 939$MeV and $m_- = 1540$MeV, they get the values
\be
  \sinh \delta = 5.5, \ \ 
  g_1 = 13.0, \ \ 
  g_2 = 3.2, \ \ 
  m_0 = 270 {\rm MeV}
  \label{eq:dkdata}
\ee
and 
\be
  |g_{\pi N_+ N_+}| = 9.9, \ \
  |g_{\pi N_- N_-}| = 16.2,\ \ 
  |g_{\pi N_+ N_-}| = 0.7
\ee
An interesting point here is that the off-diagonal coupling $g_{\pi N_+ N_-}$
is very small as compared with the other two.
This comes from the large mixing parameter $\delta$, 
i.e., $\sinh g_1 \sigma/m_0 \simeq 5.5$ and implies that the mass 
$m_0$ of the baryons in the chiral restoration phase is also small compared 
to the mass scale of ordinary baryons.

\section{Extension to $\bf SU(3)$ with the mirror fermion}
\label{sc:su3}

  In the previous section, we find that the introduction of the mirror fermion
is essential for a nontrivial coupling of the positive and negative parity
baryons.
We now apply the same formulation to the $SU(3)_L \times SU(3)_R$ chiral
symmetry.
The aim is to show whether the idea of chiral mirror baryons can be 
applied to all the octet baryons and also to see whether the quantitative
results in $SU(2)$ remain valid.

  Extension to $SU(3)$ is not straightforward since the baryon field belongs
to the adjoint representation of $SU(3)$.
It is known that the $SU(3)$ baryon in the linear realization has an ambiguity 
in specifying transformation properties under 
$SU(3)_L \times SU(3)_R$\cite{Ge84}.
The only requirement for the $SU(3)$ baryons is that they form an octet
representation of the diagonal subgroup of $SU(3)_V$.
In fact, there are three linear representations 
$(8,1)+(1,8)$, $(3,\bar{3})+(\bar{3},3)$ and $(6,3)+(3,6)$ which satisfy 
this requirement\cite{CJ97}.
Instead, here we construct a chiral effective model in the nonlinear 
representation.
By taking this representation, the baryon octet transforms as an adjoint
representation under $SU(3)_V$ which is the symmetry after SCSB and the 
meson-baryon couplings form derivative couplings.
This also enables us to perform a chiral perturbation expansion.

  First we rewrite the $SU(2)$ linear sigma model with a mirror fermion 
introduced in the previous section into a nonlinear form.
The Lagrangian (\ref{eq:su2lag3}) is written as
\bea
  {\cal L} &=& \psb_+ i \pa \sla \psi_+ + \psb_- i \pa \sla \psi_- 
  -f(g_1-g_2)(\psb_{+R} M^\dagger \psi_{+L} + \psb_{+L} M \psi_{+R}) 
  \nonumber \\
  && -f(g_1+g_2)(\psb_{-R} M \psi_{-L} + \psb_{-L} M^\dagger \psi_{-R})
  \nonumber \\
  && +m_0 (\psb_{+R}\psi_{-L}-\psb_{+L}\psi_{-R}-\psb_{-R}\psi_{+L}
   +\psb_{-L}\psi_{+R})
  \label{eq:su2lag4}
\eea
where $M=(\sigma+ i \vec{\pi}\cdot\vec{\tau})/f$ and ${\cal L}_M$ 
is omitted.
The chiral field $M$ transforms under $SU(2)_L \times SU(2)_R$ as
\be
  M \lto M' = L M R^\dagger
\ee
In order to rewrite eq.(\ref{eq:su2lag4}) into a nonlinear representation,
we redefine the NG-boson fields in the exponential parametrization,
\be
  \xi = e^{i \pi / f}
\ee
with $\pi = \vec{\pi}\cdot \vec{\tau} / 2$.
$\xi$ is related to the linear form $M$ by
\be
  \xi^2 = M
\ee
and transforms under $SU(2)_L \times SU(2)_R$ as
\be
  \xi \lto \xi' = L \xi U^\dagger = U \xi R^\dagger
  \label{eq:xitr}
\ee
Here the matrix $U$ has been defined and is a nonlinear function of 
$L$, $R$ and $\xi$.
The role of $U$ is to preserve the decomposition $M=\xi^2$.
This is verified as
\be
  M \lto M' = L M R^\dagger = L \xi \xi R^\dagger = L \xi U^\dagger
  U \xi R^\dagger
\ee
Eq.(\ref{eq:xitr}) ensures that $M' = LMR^\dagger=\xi'^2$.

We now rewrite the baryon fields $\Psi$ in terms of $\chi$ and $\xi$
\be
  (\psi_{+L}, \psi_{-R}) = \xi (\chi_{+L},\chi_{-R}), \ \ \ \ \  
  (\psi_{+R}, \psi_{-L}) = \xi^\dagger (\chi_{+R},\chi_{-L})
\ee
The field $\chi$ transforms nonlinearly under $SU(2)_L \times SU(2)_R$
\be
  \chi \lto \chi' = U \chi
\ee
where it  reduces to 
\be
  \chi \lto \chi' = V \chi, \ \ (L=R=V)
\ee
for the diagonal vector transformation under $SU(2)_V$.

Now the Lagrangian (\ref{eq:su2lag4}) turns into the following form:
\bea
  {\cal L} &=& \chb i \pa \sla \chi  + i  \chb \gamma^\mu V_\mu  \chi  
  + g_A \chb \rho_3 \gamma_5 \gamma^\mu A_\mu  \chi  \nonumber \\
  & & -g_1 f \chb \chi  + g_2 f \chb \rho_3 \chi 
  - i m_0  \chb \rho_2 \gamma_5 \chi 
\eea
with $V_\mu = \frac{1}{2} ( \xi^\dagger \pa_\mu \xi + \xi \pa_\mu \xi^\dagger
)$ and $A_\mu = i \frac{1}{2} ( \xi^\dagger \pa_\mu \xi - \xi \pa_\mu 
\xi^\dagger)$.
By taking the nonlinear representation, we have introduced the axial-vector 
coupling constant $g_A$ as a free parameter.

  Once a nonlinear effective Lagrangian of $SU(2)$ has been constructed,
an extension to the $SU(3)$ case is straightforward with a few
modifications.
In $SU(3)$, the baryon fields, $\chi$, are octet representations and 
transform as
\be
  \chi \lto \chi' = U \chi U^\dagger
\ee
Furthermore, the axial-vector coupling constants are split into two terms 
of so-called $F$- and $D$-couplings.
Combining these two considerations, 
one gets the following form of the $SU(3)$ Lagrangian,
\bea
    {\cal L} &=& \tr ( \chb i \pa \sla \chi ) + i \tr ( \chb \gamma^\mu
  \left[ V_\mu , \chi \right] ) \nonumber \\
  & & + F \tr ( \chb \rho_3 \gamma_5 \gamma^\mu \left[ A_\mu , \chi \right] ) 
   + D \tr ( \chb \rho_3 \gamma_5 \gamma^\mu \left\{ A_\mu , \chi \right\} )  
    \nonumber \\
  & & -g_1 f \tr ( \chb \chi ) + g_2 f \tr ( \chb \rho_3 \chi )
  - i m_0 \tr ( \chb \rho_2 \gamma_5 \chi )
  \label{eq:su3lag1}
\eea
where $V_\mu$ and $A_\mu$ are defined in the same way as $SU(2)$ except that
the NG-boson field $\xi$ is in terms of the Gell-Mann matrix defined by
\be
  \xi = e^{i\pi/f},\ \ \ \ \ \pi=\pi^a \frac{\lambda^a}{2}, \quad\quad
  (a=1...8)
\ee
Note that in eq.(\ref{eq:su3lag1}) there are $\rho_3$ factors in the
$F$- and $D$-coupling terms.
This is because we take $\chi_-$ as a mirror fermion field.

  Since the mass term in eq.(\ref{eq:su3lag1}) includes an off-diagonal part 
proportional to $m_0$, it must be diagonalized.
In the same way as in the previous section, one gets the diagonalized form as 
follows,
\bea
  {\cal L} &=& \tr ( \nb  i \pa \sla N ) - \tr ( \nb m N )
  + i \tr ( \nb \gamma^\mu \left[ V_\mu , N \right] ) \nonumber \\
  &+& F \left[ \tanh \delta \, \tr ( \nb \rho_3 \gamma_5 \gamma^\mu \left[
  A_\mu , N \right] ) - \frac{1}{\cosh \delta} \tr ( \nb \rho_1 \gamma^\mu
  \left[ A_\mu , N \right] ) \right] \nonumber \\
  &+& D \left[ \tanh \delta \, \tr ( \nb \rho_3 \gamma_5 \gamma^\mu \left\{
  A_\mu , N \right\} ) - \frac{1}{\cosh \delta} \tr ( \nb \rho_1 \gamma^\mu
  \left\{A_\mu , N \right\} ) \right]
   \label{baslag}
\eea
with $\sinh \delta = g_1 f/m_0$. 
$N = \left( \begin{array}{c} N_+ \\ N_- \end{array} \right)$ 
is the nucleon field in the basis which diagonalizes the mass matrix as
\be
  m = \left( 
  \begin{array}{cc}
        m_+ = -g_2 f+ m_0 \sqrt{\left( \frac{g_1 f}{m_0} \right)^2 +1} & 0 \\
    0 & m_- = +g_2 f+ m_0 \sqrt{\left( \frac{g_1 f}{m_0} \right)^2 +1}
  \end{array} \right)
\ee
In eq.(\ref{baslag}), the parameter $\delta$ is new and does not exist in 
the conventional nonlinear chiral effective model\cite{Ge84}.
$F \, \tanh \delta$ and $D \, \tanh \delta$ correspond to the conventional
$F$ and $D$, and therefore, the axial-vector coupling constant of the
nucleon, for example,
in the tree approximation is $g_A = (F+D) \tanh \delta$.
The terms with $\rho_1$ involve Yukawa couplings between $N_+$ and
$N_-$.
If we choose $\tanh \delta = 1$, corresponding to $m_0 = 0$, then 
$N_+$ and $N_-$ completely decouple from each other because
$1/\cosh \delta=0$.
As $m_0$ increases, so does the $N_+$-$N_-$ Yukawa coupling.
The parameter $\delta$ or $g_1/m_0$ controls the strength of the 
Yukawa couplings.
We adopt this Lagrangian(\ref{baslag}), which has the parameters $F$, $D$, 
$\delta$, the positive parity baryon mass $m_+$, and the negative one $m_-$
for strong decays of negative parity baryons.

\section{Strong decays of the octet $\frac{\bf 1}{\bf 2}^-$  baryons}
\label{sc:decay}

\subsection{Formulation}  \label{ssc:form}

 Strong decays of the flavor-octet negative parity baryons  are calculated here
to leading order of chiral logarithm.
Since the baryon masses are large and we are interested in processes with 
characteristic energy smaller than the baryon masses, we use the formulation
of heavy baryon chiral perturbation theory(HBChPT)\cite{JM91a}.
The idea is to write the baryon momentum as $p^\mu = m v^\mu + k^\mu$,
where $m$ is the baryon mass and $v^\mu$ is the velocity chosen so that
the residual off-shell momentum $k^\mu$ is small as compared to the chiral 
symmetry breaking scale, $\Lambda_\chi\sim 1$GeV\cite{MG}.
The effective Lagrangian is written in terms of the baryon fields $B$ which
satisfy the positive energy condition $v \sla B = B$ and obey the Dirac 
equation $i \pa \sla B = 0$ that does not include the mass term.

  Defining a heavy baryon field $B$ as 
\be
  \left(
  \begin{array}{c}
    B_+(x) \\ B_-(x)
  \end{array}
  \right) = e^{i m_+ v \cdot x} \left(
  \begin{array}{c}
    N_+(x) \\ N_-(x)
  \end{array}
  \right)
\ee
the Lagrangian (\ref{baslag}) becomes, including meson fields,
\bea
  {\cal L} &=& \tr ( \bb  i v \cdot \pa B ) 
  + i \tr ( \bb v^\mu \left[ V_\mu , B \right] ) \nonumber \\
  &+& F \left[ \tanh \delta \, \tr ( \bb \rho_3 S^\mu \left[
  A_\mu , B \right] ) - \frac{1}{\cosh \delta} \tr ( \bb \rho_1 v^\mu
  \left[ A_\mu , B \right] ) \right] \nonumber \\
  &+& D \left[ \tanh \delta \, \tr ( \bb \rho_3 S^\mu \left\{
  A_\mu , B \right\} ) - \frac{1}{\cosh \delta} \tr ( \bb \rho_1 v^\mu
  \left\{ A_\mu , B \right\} ) \right] \nonumber \\
  &+& \Delta m \tr ( \bb_- B_- ) \nonumber \\
  &+& \frac{f^2}{4} \tr (\pa_\mu \Sigma \pa^\mu \Sigma^\dagger )
  + a \tr \{ M_q (\Sigma + \Sigma^\dagger) \}
\eea
where $\Delta m = m_- - m_+$ is the mass difference between $B_+$ and $B_-$,
and $M_q$ is the quark mass matrix, for which we assume
\be
  M_q = \left( \begin{array}{ccc}
    0 & 0 & 0   \\
    0 & 0 & 0   \\
    0 & 0 & m_s 
  \end{array} \right)
\ee
It is noted that the $\Delta m$ term is included so as to see the decay of the
negative parity baryons going into the positive ones.
The baryon fields are explicitly given by
\be
  B = \left( \begin{array}{ccc}
    \frac{1}{\sqrt{2}} \Sigma^0 + \frac{1}{\sqrt{6}} \Lambda &
    \Sigma^+ & p \\
    \Sigma^- & -\frac{1}{\sqrt{2}} \Sigma^0 + \frac{1}{\sqrt{6}} \Lambda & n \\
    \Xi^- & \Xi^0 & -\frac{2}{\sqrt{6}} \Lambda
  \end{array} \right)
\ee
and the meson fields by
\be
  \pi = \frac{1}{\sqrt{2}} \left( \begin{array}{ccc}
    \frac{1}{\sqrt{2}} \pi^0 + \frac{1}{\sqrt{6}} \eta & \pi^+ & K^+ \\
    \pi^- & -\frac{1}{\sqrt{2}} \pi^0 + \frac{1}{\sqrt{6}} \eta & K^0 \\
    K^- & \bar{K^0} & -\frac{2}{\sqrt{6}} \eta
  \end{array} \right)
\ee

  We may assign the $\frac{1}{2}^-$ octet baryons to the observed baryons such 
as  $N(1535)$, $\Lambda(1670)$ and $\Sigma(1750)$\cite{PDG}.
The $\Lambda (1405)$, which is the lightest negative parity baryon in the
$\Lambda$ sector, is associated mainly with the flavor singlet state\cite{PDG}.
As for $\Sigma$, another possible choice is $\Sigma(1630)$, but the
experimental data for $\Sigma(1630)$ decays are so poor that we here 
compare our results with the well-established $\Sigma(1750)$. 
The $\Xi$ sector has little been studied experimentally.
The QCD sum rule suggests $\Xi(1690)$ to be one of the members of the 
$\frac{1}{2}^-$ octet baryons.
But the decays of $\Xi(1690)$ are little known qualitatively, so they
are not treated here. 

  In the calculation of the axial-vector coupling constants of the octet 
baryons,
the coupling constants $F$ and $D$ change if the couplings of the decuplet 
baryons are taken into account\cite{JM91b}.
We, however, have not included them because 
as will be seen later, decays of the negative parity baryons depend more on
the parameter $\delta$ sensitively than on $F$ and $D$. 

\subsection{Calculation}  \label{ssc:calc}

  We calculate the decay widths of the following processes:
$N(1535) \to N + \pi$, $N + \eta$, 
$\Lambda(1670) \to N + \bar{K}$, $\Sigma + \pi$,
$\Lambda + \eta$ and $\Sigma(1750) \to N + \bar{K}$,
$\Lambda + \pi$, $\Sigma + \pi$, $\Sigma + \eta$.  
The $\eta$ meson here is treated as the pure octet component, $\eta_8$, of
$SU(3)_V$.
The mixing of the singlet $\eta_1$ with $\eta_8$ remains to be considered in 
the future.

  A decay width is given by
\be
 \Gamma = \frac{\Delta m^2}{8 \pi f^2}\sqrt{\Delta m^2 - M^2} |{\cal F}|^2
\ee
where $M$ is the mass of the final state meson and the invariant amplitude 
$\cal F$ is calculated up to the order of the chiral logarithm $O(M \log M)$.
The relevant diagrams are shown in Figs.1 and 2.
The amplitudes for each process are given as follows.
\bea
  {\cal F} 
  &=& \alpha \left[ 1 + \frac{1}{16 \pi^2 f^2} \left\{
    \lambda_1 W(M_K) + \lambda_2 W(M_\eta) - \lambda_3 G 
    + \lambda_4 H \right\} \right] \nonumber \\
  &+& \frac{1}{16 \pi^2 f^2} \left\{ \beta_1 G + \beta_2 X(M_K) + \beta_3
    X(M_\eta) + \beta_4 H \right\} 
  \label{amplitude}
\eea 
where $M_K$ and $M_\eta$ are the masses of $K$ and $\eta$, respectively.
In the amplitude (\ref{amplitude}), these terms are classified as follows;
the terms of $\alpha$ are from tree diagrams, the terms of $\lambda_i$
are from the wave function renormalization, and the terms of $\beta_i$
are from one-loop corrections (see Figs.1 and 2).
The coefficients $\alpha$, $\lambda_i$ and $\beta_i$ depend on $F$, $D$ 
and $\delta$, and are given for each decay separately. 
For the process of $p_- \to p + \pi^0$, for example, they are given by
\bea
 \alpha &=& \frac{1}{2}(D + F) \frac{1}{\cosh \delta} \\
 \lambda_1 &=& \left( \frac{10}{6} D^2 - 2 DF + 3 F^2 \right)
   \frac{1}{\cosh^2 \delta} \nonumber \\ 
 \lambda_2 &=& \left( \frac{1}{6} D^2 - DF + \frac{3}{2} F^2 
   \right) \frac{1}{\cosh^2 \delta} \nonumber \\ 
 \lambda_3 &=& \left( \frac{17}{9} D^2 - \frac{10}{3} DF + 5 F^2
   \right) \tanh^2 \delta - \frac{1}{6} \nonumber \\ 
 \lambda_4 &=& \left( \frac{3}{2} D^2 + 3 DF + \frac{3}{2} F^2
   \right) \frac{1}{\cosh^2 \delta} \\
 \beta_1 &=& \left( \frac{1}{3} D^3 + \frac{1}{3} D^2F + DF^2 - 3 F^3 \right)
   \frac{\tanh^2 \delta}{\cosh \delta}
   -\frac{1}{12} ( D + F ) \frac{1}{\cosh \delta} \nonumber \\
 \beta_2 &=& \left( -\frac{1}{6} D^3 + \frac{1}{6} D^2F - \frac{1}{2} DF^2
   + \frac{1}{2} F^3 \right) \frac{1}{\cosh^3 \delta} \nonumber \\
 \beta_3 &=& \left( \frac{1}{24} D^3 - \frac{5}{24} D^2F + \frac{1}{8} DF^2
   + \frac{3}{8} F^3 \right) \frac{1}{\cosh^3 \delta} \nonumber \\
 \beta_4 &=& \left( -\frac{1}{4} D^3 - \frac{3}{4} D^2F - \frac{3}{4} DF^2
   - \frac{1}{4} F^3 \right) \frac{1}{\cosh^3 \delta}
\eea
$G$, $H$, $W$ and $X$ contain loop integrals and are given by
\be
  G = M_K^2 \log \frac{M_K^2}{\mu^2}
\ee
\be
  H = 3 \Delta m^2 \log \frac{4 \Delta m^2}{\mu^2}
\ee
\bea 
  W(M) &=& ( -\frac{M^2}{2} + 3 \Delta m^2) \log \frac{M^2}{\mu^2} \nonumber \\
       & & - \Delta m \left(
           \frac{\Delta m^2}{\sqrt{\Delta m^2 - M^2}} + 2 
           \sqrt{\Delta m^2 - M^2} \right) \log 
           \frac{\Delta m - \sqrt{\Delta m^2 - M^2}}
                {\Delta m + \sqrt{\Delta m^2 - M^2}}
\eea
\be
  X(M) = ( M^2 - 2 \Delta m^2) \log \frac{M^2}{\mu^2} 
         + 2 \Delta m \sqrt{\Delta m^2 - M^2} \log
           \frac{\Delta m - \sqrt{\Delta m^2 - M^2}}
                {\Delta m + \sqrt{\Delta m^2 - M^2}}
\ee  
We calculate the loop integrals taking into account the mass difference 
between $m_+$ and $m_-$, i.e. $\Delta m$.
More details of computation are explained in Appendix A.
The coefficients $\alpha$, $\lambda_i$ and $\beta_i$ used for the $\Lambda_-$
and $\Sigma_-$ decays are given in Appendix B.

  It should be noted that the tree contribution $\alpha$ is purely $SU(3)$
symmetric and is determined completely by the $F/D$ ratio.
All the $SU(3)$ breaking effects are introduced by the chiral log terms
where the meson mass and baryon mass differences are taken into account.

\subsection{Results}  \label{ssc:result}

  We show decay widths of various processes as function of $\tanh \delta$  
in Fig.3 to Fig.11.
Here we have fixed $F \tanh \delta = 0.56$ and $D \tanh \delta = 0.33$,
which are determined so as to fit the axial-vector coupling constants
of the octet baryons up to the chiral logarithm in \cite{JM91a}.
These numbers are slightly modified when adjusted 
at the tree level; $(F \tanh \delta)_{\rm tree} = 0.80$ and
$(D \tanh \delta)_{\rm tree} = 0.50$.
The qualitative behaviors of the following results do not depend on the choice 
the values of $F \tanh \delta$ and $D \tanh \delta$,
but are rather sensitive to the value of $\delta$.

  We find that our results prefer a large value of $\tanh \delta$, 
or a small $m_0$.
Especially the small decay widths of $N_- \to N\pi$, $\Lambda_- \to N\bar{K}$
and $\Lambda_- \to \Sigma\pi$ require $\tanh \delta\simeq 0.98$ or larger.
The corresponding values of $m_0$, defined by
\be
  m_0 = \frac{m_+ + m_-}{2} \frac{1}{\cosh \delta}
\ee
are less than $250$MeV as shown in Fig.12.
When taking into account large uncertainties in experimental data,
we conclude
$0.91 \lsim \tanh \delta \lsim 0.99$ and $200 \lsim m_0 \lsim 500$MeV.
The value $m_0$ can be fixed less precisely than $\tanh \delta$ 
because of a rapid change of $m_0$ in this region as shown in Fig.12.
These observations agree with the analysis of
the $N_- \to N_+ + \pi$ decay by DK\cite{DK89}(See eq.(\ref{eq:dkdata})).
We also see in the figures that the chiral log corrections (solid lines) 
to the $SU(3)$ limit (dashed lines) are small for the decays of $N(1535)$,
but substantial for those of $\Lambda(1670)$ and $\Sigma(1750)$.
Therefore we may conclude that the $SU(3)$ breaking effect is larger for the 
decays of strange baryons.

  The predictions for two processes of $N_- \to N_+ + \eta$(Fig.4) and
$\Sigma_- \to \Sigma_+ + \eta$(Fig.11) are somewhat smaller than the
experiment.
One of the motivations of the present study is to see whether the relatively
large decay widths of these two processes accompanied by $\eta$ are explained
or not.
However, within the present approach which is based on $SU(3)$
symmetry, it seems difficult to reproduce this property, that is, 
when parameters are fixed using the small decay rate of $N(1535)\to N+\pi$,
the $SU(3)$ symmetry leads to a small decay rate $B_- \to B_+ + \eta$ also.
We need more theoretical insight as well as more accurate experimental data
to understand physics of resonance decays.
 
\section{Discussions} \label{sc:discuss}

  In the previous section, we have studied strong decays of negative parity
baryons within the parity pair model based on the mirror fermion 
formulation \`a la DeTar and Kunihiro\cite{DK89}.
However, as is shown in Sec.\ref{sc:su2}, 
this is not a unique choice of introducing the parity pair baryons.
For instance, the other choice, ``naive" model, is possible, although it 
gives a trivial result for the baryon decay.
Namely, in the tree level, the negative parity baryons do not decay into 
positive parity baryons and a NG-boson.
The small decay rate of, for instance, $N(1535) \to N + \pi$ could therefore
be consistent with both models.

How are they different from each other?
The ``mirror" model contains a new parameter $m_0$, which allows a finite
mass of the baryons in the limit of chiral restoration,
whereas the ``naive" model forces the masses of the baryons to vanish in 
this limit.
Therefore, if the parameter $m_0$ is large, two approaches would be 
distinguished most drastically.
However, we have found in the previous section that this parameter $m_0$
is small, say $m_0 \lsim 500$MeV, or even smaller $m_0 \sim 200$MeV 
when small decay rates of negative parity baryons are adopted.
Thus the ``mirror" model is not so different from the naive model in this
regard.

  However, there is another distinctive feature between the two 
approaches, that is the relative sign of the axial-vector coupling 
constant $g_A$.
The sign of $g_A$ of the negative parity baryon is opposite to that of the 
positive one in the mirror model\cite{DK89}.
Although the direct measurement of $g_A$ of the negative parity baryon
seems difficult,
$g_A$ is related to the coupling constant of the pion (NG-boson) through the
Goldberger-Treiman relation in the soft-pion limit.
Therefore, the sign of the pion exchange interaction between the positive 
and negative parity baryons is directly connected to the relative sign of 
$g_A$.
It may be extremely interesting if relative sign of 
$g_A$ is studied both theoretically and experimentally.

  Finally we comment on the analysis in the correlation function approach.
In \cite{JO96b}, it was found that the correlation function for the $\pi N_+ 
\, N_-$ coupling vanishes in the chiral and soft-pion limit.
Their correlation function happens to correspond to the naive assignment of
the parity pair nucleons.
Therefore the results of \cite{JO96b} agree with the present analysis both
for the coupling constant and the masses of the baryons\cite{JO96a} when
the chiral symmetry is restored.

\section{Summary and conclusion} \label{sc:concl}

  We have studied the decays of $\frac{1}{2}^-$ octet baryons in the
$SU(3)_L \times SU(3)_R$ chiral effective model.
The model consists of a parity pair of baryons \`a la DeTar and Kunihiro
where the positive and negative parity baryons behave like mirror
fermions.
This model contains a new mass parameter $m_0$ which determines the 
degenerate mass of the positive and negative parity baryons when the chiral
symmetry is restored and also fixes the strength of the coupling of the 
both parity baryons with chiral mesons.

  Strong $\pi$, $K$ and $\eta$ decays of $N_-$, $\Lambda_-$ and
$\Sigma_-$ have been calculated up to the order of the chiral logarithm  
in the chiral perturbation theory.
The baryons and the baryon resonances have been treated as heavy baryons.
We have found that the results agree qualitatively with the present experiment 
with the parameter $\tanh \delta \gsim 0.9$ or $m_0 \lsim 500$MeV.
This is consistent with the previous result for the $SU(2)_L \times
SU(2)_R$ linear sigma model\cite{DK89}.
The chiral log contributions, which incorporate $SU(3)$ breaking effects,
are found to be significant in hyperon decays, 
but do not affect the qualitative features of the tree calculation.

  As the present analysis prefers a small $m_0$,
baryon masses in the limit of chiral restoration may not distinguish
the ``mirror" from the ``naive" chiral assignment.
However,  the ``mirror" model predicts the opposite signs for the 
axial-vector coupling constants $g_A$ for the positive and negative 
parity baryons.
Therefore it would be extremely interesting if the 
relative sign of the pion-baryon couplings will be determined experimentally.

  Present experimental data are not accurate enough to impose a firm 
constraint on the parameter $\tanh \delta$ or $m_0$.
More accurate experimental data are desired in order to study further the
structure of baryons and baryon excitations. \\

{\Large \bf Appendices} 
\appendix
\section{An example of one-loop calculation}

  Basically we follow the evaluation of the conventional method of 
HBChPT\cite{JM91a}, but we take into account the mass difference between the 
negative and positive parity baryons, $\Delta m$.
There appear additional terms in this case. 
In order to demonstrate the new terms,
we consider a $B_- \to B_++$meson decay diagram shown in Fig.\ref{fig:app} 
as an example.
In the figure, we assume the initial and final baryons are on the mass shell
so that the momentum of the final baryon $B_+$ is $m_+ v_\mu$ and that of the
initial baryon $B_-$ is $m_- v_\mu \equiv m_+ v_\mu + p_\mu$.
Multiplying $v_\mu$ to both sides, we find $p \cdot v = m_- - m_+ = \Delta m$.

  The corresponding one-loop integral in $d$-dimensional space time is
\be
  \Sigma(v \cdot p) = 
  \mu^{4-d} i (-v \cdot p) \int \frac{d^d l}{(2\pi)^d} \frac{(v \cdot l)^2}
  {(v \cdot l - \Delta m)(v \cdot (p + l))(l^2 - M^2)}
  \label{eq:app1}
\ee
or
\bea
  \Sigma(\Delta m) &=&
  \frac{\mu^{4-d}}{i} \Delta m v^\alpha v^\beta \int \frac{d^d l}{(2\pi)^d}
  \frac{l_\alpha l_\beta}{(v \cdot l - \Delta m)(v \cdot l + \Delta m)
  (l^2 - M^2)} \nonumber \\
  &\rightarrow& \frac{\Delta m}{16 \pi^2} \left[ (-M^2 + 2 \Delta m^2)
  \log \frac{M^2}{\mu^2} - 2 \Delta m \sqrt{\Delta m^2 - M^2} \log
  \frac{\Delta m - \sqrt{\Delta m^2 - M ^2}}
  {\Delta m+ \sqrt{\Delta m^2 - M ^2}}
  \right]  \nonumber \\
  && 
\eea
at the order of the chiral logarithm after the dimensional regularization.
Other diagrams are similarly evaluated.  

  In this treatment, we have neglected the $SU(3)$ breaking effects,
but it is also possible to include the mass differences among  the octet 
baryons.   
Although these mass differences are not included in the Lagrangian.
one can incorporate the baryon masses of the intermediate states in the 
one-loop integrals in a more phenomenological way.
If we take mass differences relative to the final baryon as shown in 
Fig.\ref{fig:app2}, eq.(\ref{eq:app1}) is replaced by
\bea
  \Sigma(v \cdot p) &=&
  \mu^{4-d} i (-v \cdot p) \int \frac{d^d l}{(2\pi)^d} \frac{(v \cdot l)^2}
  {(v \cdot l - \Delta m_3)(v \cdot (p + l) - \Delta m_2)(l^2 - M^2)}
  \nonumber \\
  &=& \frac{\mu^{4-d}}{i} \Delta m_1 v^\alpha v^\beta \int 
  \frac{d^d l}{(2\pi)^d}
  \frac{l_\alpha l_\beta}{(v \cdot l - \Delta m_3)(v \cdot l + \Delta m_1
  - \Delta m_2) (l^2 - M^2)}
\eea
This equation is more complicated than eq.(\ref{eq:app1}) since the integral
calculation is needed in both $\Delta m_i > M$ and $\Delta m_i < M$ cases.
It is easy to see that our results are
little modified at $\tanh \delta \simeq 1$ since the dominant contributions in
this region are the tree diagrams and the wave function renormalization 
diagrams which have a positive parity baryon as the intermediate state.
 
\section{Coefficients used in the $\bf \Lambda_-$ and $\bf \Sigma_-$ decays}
\begin{itemize}
\item{$p_- \lto p + \eta$}
\bea
  \alpha &=& - \frac{1}{2 \sqrt 3}(D -3F) \frac{1}{\cosh \delta} \\
  \lambda_1 &=& \left( \frac{10}{6} D^2 - 2 DF + 3 F^2 \right)
    \frac{1}{\cosh^2 \delta}  \nonumber \\ 
  \lambda_2 &=& \left( \frac{1}{6} D^2 - DF + \frac{3}{2} F^2 
    \right) \frac{1}{\cosh^2 \delta} \nonumber \\
  \lambda_3 &=& \left( \frac{17}{9} D^2 - \frac{10}{3} DF + 5 F^2
    \right) \tanh^2 \delta  - \frac{1}{2} \nonumber \\
  \lambda_4 &=& \left( \frac{3}{2} D^2 + 3 DF + \frac{3}{2} F^2
    \right) \frac{1}{\cosh^2 \delta}  \\
  \beta_1 &=& \sqrt 3 \left( -\frac{11}{18} D^3 + \frac{3}{2} D^2 F 
    + \frac{3}{2} D F^2 - \frac{3}{2} F^3 \right) 
    \frac{\tanh^2 \delta}{\cosh \delta} 
    + \sqrt 3 \left( \frac{1}{12} D - \frac{1}{4} F \right)
    \frac{1}{\cosh \delta} \nonumber \\
  \beta_2 &=& \sqrt 3 \left( \frac{2}{9} D^3 - \frac{2}{3} D^2 F \right)
    \frac{1}{\cosh^3 \delta} \nonumber \\
  \beta_3 &=& \sqrt 3 \left( -\frac{1}{72} D^3 + \frac{1}{8} D^2 F
    -\frac{3}{8} D F^2 + \frac{3}{8} F^3 \right) \frac{1}{\cosh^3 \delta}
    \nonumber \\
  \beta_4 &=& \sqrt 3 \left( -\frac{1}{3} D^3 + \frac{1}{3} D^2 F   
    + \frac{5}{3} D F^2 + F^3 \right) \frac{1}{\cosh^3 \delta} 
\eea 
 
\item{$\Lambda_- \lto p + K^-$} 
\bea
  \alpha &=& -\frac{1}{2\sqrt 3} (D + 3F) \frac{1}{\cosh \delta}  \\
  \lambda_1 &=& \left( \frac{7}{6} D^2 - D F + \frac{9}{2} F^2 \right) 
    \frac{1}{\cosh^2 \delta}  \nonumber \\ 
  \lambda_2 &=& \left( \frac{5}{12} D^2 - \frac{1}{2} D F + \frac{3}{4} F^2
    \right) \frac{1}{\cosh^2 \delta} \nonumber \\
  \lambda_3 &=& \left( \frac{31}{18} D^2 - \frac{5}{3} D F + \frac{11}{2} F^2
    \right) \tanh^2 \delta  -1 \nonumber \\
  \lambda_4 &=& \left( \frac{7}{4} D^2 + \frac{3}{2} D F + \frac{3}{4} F^2
    \right) \frac{1}{\cosh^2 \delta} \\
  \beta_1 &=& \sqrt 3 \left( \frac{19}{36} D^3 - \frac{5}{8} D^2 F 
     - \frac{7}{8} D F^2 + \frac{9}{8} F^3 \right) 
     \frac{\tanh^2 \delta}{\cosh \delta} 
     + \sqrt 3 \left( \frac{7}{216} D + \frac{95}{648} F \right)
     \frac{1}{\cosh \delta} \nonumber \\
  \beta_2 &=& \sqrt 3 \left( -\frac{5}{36} D^3 +\frac{5}{12} D^2 F
     +\frac{1}{4} D F^2 - \frac{3}{4} F^3 \right) \frac{1}{\cosh^3 \delta}
     \nonumber \\
  \beta_3 &=& \sqrt 3 \left( -\frac{1}{36} D^3 + \frac{1}{4} D F^2 \right)
     \frac{1}{\cosh^3 \delta} \nonumber \\
  \beta_4 &=& \sqrt 3 \left( \frac{1}{2} D^3 - \frac{1}{2} D F^2 \right)
     \frac{1}{\cosh^3 \delta} 
\eea

\item{$\Lambda_- \lto \Sigma^0 + \pi^0$}
\bea
  \alpha &=& \frac{1}{\sqrt 3} D \frac{1}{\cosh \delta} \\
  \lambda_1 &=& \left( \frac{4}{3} D^2 + 4 F^2 \right)\frac{1}{\cosh^2 \delta}
              \nonumber \\
  \lambda_2 &=& \frac{2}{3} D^2 \frac{1}{\cosh^2 \delta}  \nonumber \\
  \lambda_3 &=& \left( \frac{20}{9} D^2 + 4 F^2 \right) \tanh^2 \delta
              - \frac{1}{6} \nonumber \\
  \lambda_4 &=& \left( \frac{4}{3} D^2 + 2 F^2 \right) \frac{1}{\cosh^2 \delta}
                \\
  \beta_1 &=& \sqrt{3} \left( \frac{17}{18} D^3 - \frac{1}{2} D F^2 \right)
    \frac{\tanh^2 \delta}{\cosh \delta} 
    - \frac{\sqrt 3}{18} D \frac{1}{\cosh \delta} \nonumber \\
  \beta_2 &=& \sqrt{3} \left( -\frac{1}{6} D^3 + \frac{1}{6} F^3 \right)
    \frac{1}{\cosh^3 \delta} \nonumber \\
  \beta_3 &=& - \frac{\sqrt 3}{9} D^3 \frac{1}{\cosh^3 \delta} \nonumber \\
  \beta_4 &=& \sqrt{3} \left( \frac{2}{9} D^3 - \frac{4}{3} D F^2 \right)
    \frac{1}{\cosh^3 \delta} 
\eea 

\item{$\Lambda_- \lto \Lambda + \eta$}
\bea
  \alpha &=& -\frac{1}{\sqrt 3} D \frac{1}{\cosh \delta} \\
  \lambda_1 &=& \left( \frac{2}{3} D^2 + 6 F^2 \right) \frac{1}{\cosh^2 \delta}
    \nonumber \\
  \lambda_2 &=& \frac{2}{3} D^2 \frac{1}{\cosh^2 \delta}  \nonumber \\
  \lambda_3 &=& \left( \frac{14}{9} D^2 + 6 F^2 \right) \tanh^2 \delta  
    - \frac{1}{2} \nonumber \\
  \lambda_4 &=& 2 D^2 \frac{1}{\cosh^2 \delta}  \\
  \beta_1 &=& \sqrt{3} \left( \frac{11}{18} D^3 - \frac{3}{2} D F^2 \right)
            \frac{\tanh^2 \delta}{\cosh \delta} 
            +\sqrt{3} \left( \frac{37}{216} D + \frac{1}{216} F \right)
            \frac{1}{\cosh \delta} \nonumber \\
  \beta_2 &=& \sqrt{3} \left( -\frac{1}{18} D^3 + \frac{1}{2} D F^2 \right)
            \frac{1}{\cosh^3 \delta} \nonumber \\
  \beta_3 &=& - \frac{\sqrt 3}{9} D^3 \frac{1}{\cosh^3 \delta} \nonumber \\
  \beta_4 &=& \frac{2\sqrt 3}{3} D^3 \frac{1}{\cosh^3 \delta} 
\eea

\item{$\Sigma_-^0 \lto p + K^-$}
\bea
  \alpha &=& \frac{1}{2} ( D - F ) \frac{1}{\cosh \delta}\\
  \lambda_1 &=& \left( \frac{11}{6} D^2 - D F + \frac{5}{2} F^2 \right)
    \frac{1}{\cosh^2 \delta}  \nonumber \\
  \lambda_2 &=& \left( \frac{5}{12} D^2 - \frac{1}{2} D F + \frac{3}{4} F^2 
    \right) \frac{1}{\cosh^2 \delta}  \nonumber \\
  \lambda_3 &=& \left( \frac{43}{18} D^2 - \frac{5}{3} D F + \frac{7}{2} F^2
    \right) \tanh^2 \delta -1 \nonumber \\
  \lambda_4 &=& \left( \frac{13}{12} D^2 + \frac{3}{2} D F + \frac{11}{4} F^2
    \right) \frac{1}{\cosh^2 \delta} \\
  \beta_1 &=& \left( \frac{7}{12} D^3 - \frac{13}{12} D^2 F + \frac{7}{6} DF^2
            + \frac{3}{4} F^3 \right) \frac{\tanh^2 \delta}{\cosh \delta}
            +\left( -\frac{1}{24} D + \frac{1}{24} F \right)
            \frac{1}{\cosh \delta} \nonumber \\
  \beta_2 &=& \left( -\frac{1}{12} D^3 - \frac{1}{12} D^2 F - \frac{1}{4} D F^2
            - \frac{1}{4} F^3 \right) \frac{1}{\cosh^3 \delta} \nonumber \\
  \beta_3 &=& \left( -\frac{1}{12} D^3 + \frac{1}{3} D^2 F - \frac{1}{4} D F^2
            \right) \frac{1}{\cosh^3 \delta} \nonumber \\
  \beta_4 &=& \left( -\frac{1}{6} D^3 - \frac{1}{6} D^2 F - \frac{1}{2} D F^2
            - \frac{1}{2} F^3 \right) \frac{1}{\cosh^3 \delta}
\eea

\item{$\Sigma_-^0 \lto \Lambda + \pi^0$}
\bea
  \alpha &=& \frac{1}{\sqrt 3} D \frac{1}{\cosh \delta} \\
  \lambda_1 &=& \left( \frac{4}{3} D^2 + 4 F^2 \right)\frac{1}{\cosh^2 \delta}
    \nonumber \\
  \lambda_2 &=& \frac{2}{3} D^2 \frac{1}{\cosh^2 \delta}  \nonumber \\
  \lambda_3 &=& \left( \frac{20}{9} D^2 + 4 F^2 \right) \tanh^2 \delta
    - \frac{1}{6} \nonumber \\
  \lambda_4 &=& \left( \frac{4}{3} D^2 + 2 F^2 \right) \frac{1}{\cosh^2 \delta}
                \\
  \beta_1 &=& \sqrt 3 \left( \frac{17}{18} D^3 - \frac{1}{2} D F^2 \right)
            \frac{\tanh^2 \delta}{\cosh \delta} 
            +\sqrt 3 \left( \frac{7}{144} D -\frac{1}{1296} F \right)
            \frac{1}{\cosh \delta} \nonumber \\
  \beta_2 &=& \sqrt 3 \left( - \frac{1}{6} D^3 + \frac{1}{6} D^2 F \right)
            \frac{1}{\cosh^3 \delta} \nonumber \\
  \beta_3 &=& -\frac{\sqrt 3}{9} D^3 \frac{1}{\cosh^3 \delta} \nonumber \\
  \beta_4 &=& \sqrt 3 \left( \frac{2}{9} D^3 - \frac{4}{3} D F^2 \right)
            \frac{1}{\cosh^3 \delta} 
\eea

\item{$\Sigma_-^0 \lto \Sigma^- + \pi^+$}
\bea
  \alpha &=& F \frac{1}{\cosh \delta} \\
  \lambda_1 &=& ( 2 D^2 + 2 F^2 ) \frac{1}{\cosh^2 \delta} \nonumber \\
  \lambda_2 &=& \frac{2}{3} D^2 \frac{1}{\cosh^2 \delta} \nonumber \\
  \lambda_3 &=& \left( \frac{26}{9} D^2 + 2 F^2 \right) \tanh^2 \delta 
              - \frac{1}{4} \nonumber \\
  \lambda_4 &=& \left( \frac{2}{3} D^2 + 4 F^2 \right) 
              \frac{1}{\cosh^2 \delta} \\
  \beta_1 &=& \left( \frac{5}{9} D^2 F - F^3 \right) 
            \frac{\tanh^2 \delta}{\cosh \delta} 
            +\left( -\frac{37}{144} D -\frac{37}{144} F \right)
            \frac{1}{\cosh \delta} \nonumber \\
  \beta_2 &=& \left( -\frac{1}{2} D^2 F + \frac{1}{2} F^3 \right) 
            \frac{1}{\cosh^3 \delta} \nonumber \\
  \beta_3 &=& \frac{1}{3} D^2 F \frac{1}{\cosh^3 \delta} \nonumber \\
  \beta_4 &=& -\frac{2}{3} D^2 F \frac{1}{\cosh^3 \delta} 
\eea

\item{$\Sigma_-^0 \lto \Sigma^0 + \eta$}
\bea
  \alpha &=& \frac{1}{\sqrt 3} D \frac{1}{\cosh \delta} \\
  \lambda_1 &=& ( 2 D^2 + 2 F^2 ) \frac{1}{\cosh^2 \delta} \nonumber \\
  \lambda_2 &=& \frac{2}{3} D^2 \frac{1}{\cosh^2 \delta} \nonumber \\
  \lambda_3 &=& \left( \frac{26}{9} D^2 + 2 F^2 \right) \tanh^2 \delta
              - \frac{1}{2} \nonumber \\
  \lambda_4 &=& \left( \frac{2}{3} D^2 + 4 F^2 \right) 
              \frac{1}{\cosh^2 \delta} \\
  \beta_1 &=& \frac{\sqrt 3}{18} D^3 \frac{\tanh^2 \delta}{\cosh \delta} 
            -\frac{\sqrt 3}{6} D \frac{1}{\cosh \delta} \nonumber \\
  \beta_2 &=& \sqrt 3 \left( -\frac{1}{6} D^3 - \frac{7}{6} D F^2 \right)
            \frac{1}{\cosh^3 \delta} \nonumber \\
  \beta_3 &=& \frac{\sqrt 3}{9} D^3 \frac{1}{\cosh^3 \delta} \nonumber \\
  \beta_4 &=& \sqrt 3 \left( -\frac{2}{9} D^3 + \frac{4}{3} D F^2 \right)
            \frac{1}{\cosh^3 \delta} 
\eea
\end{itemize}

\clearpage

\begin{figure}[t]
	\hspace*{1cm}
	\epsfxsize=380pt
	\epsfbox{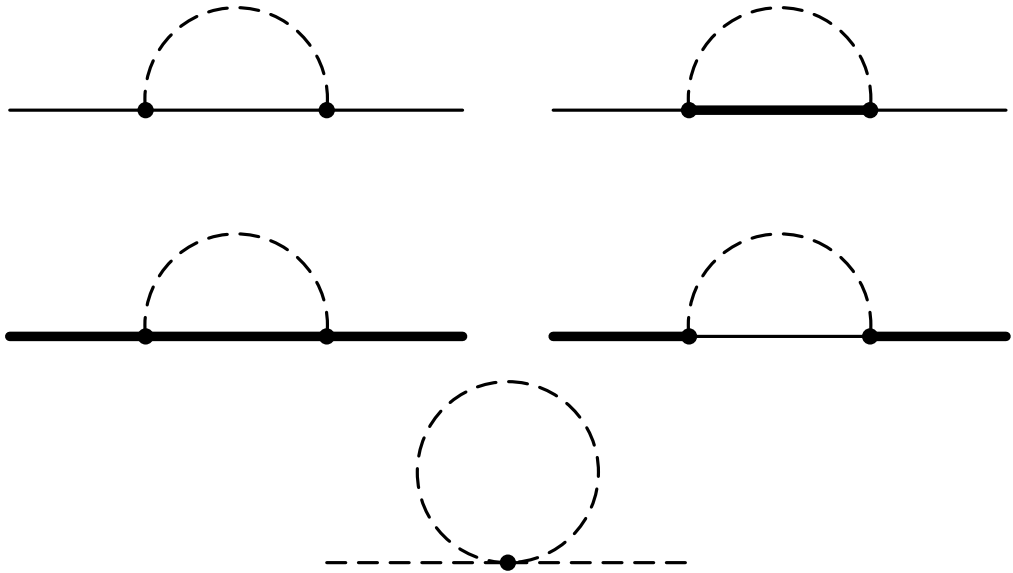}
	\caption{One-loop diagrams contributing to the wave function 
                 renormalization.
	         Thick lines denote negative parity baryons, 
                 thin lines positive ones, and dashed lines mesons.}
\end{figure}

\begin{figure}[h]
	\hspace*{1cm}
	\epsfxsize=380pt
	\epsfbox{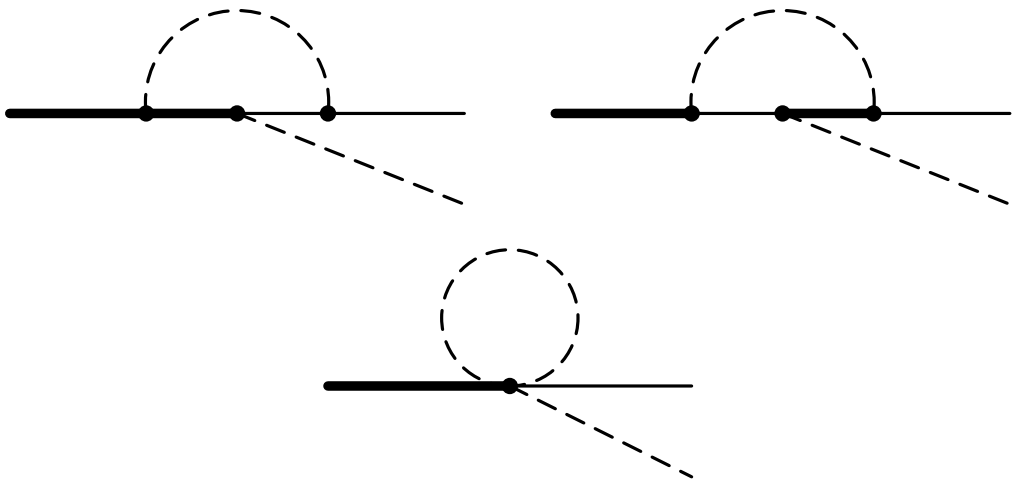}
	\caption{Diagrams contributing to the one-loop renormalization}
\end{figure}

\begin{figure}[t]
	\hspace*{3cm}
        \epsfxsize=310pt
        \epsfbox{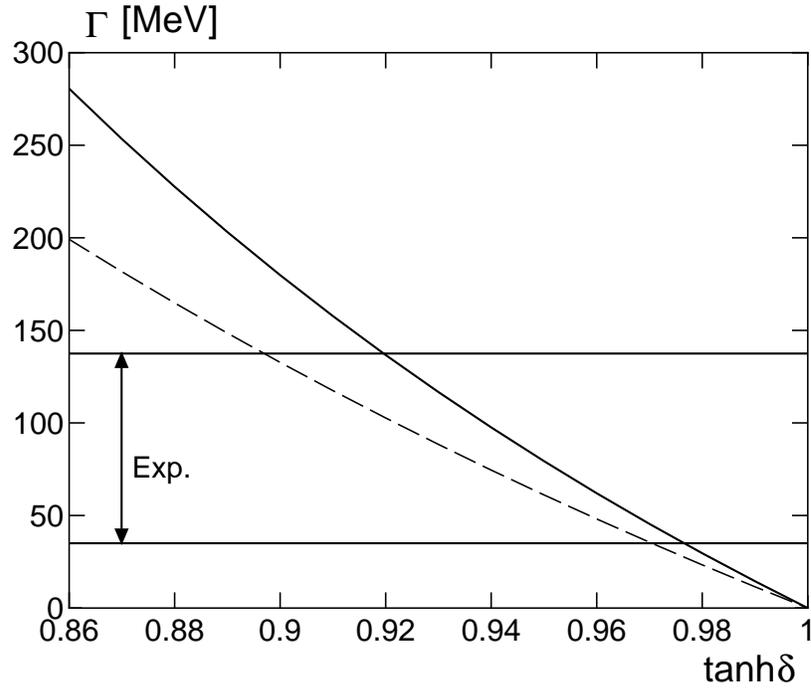}
        \caption{Decay widths of the negative parity nucleon, $N_- \to N \pi$
          The solid line includes the one-loop corrections and the dashed line
          the tree level only. }
        \label{fig:nnpi}
\end{figure}

\begin{figure}[h]
	\hspace*{3cm}
        \epsfxsize=310pt
        \epsfbox{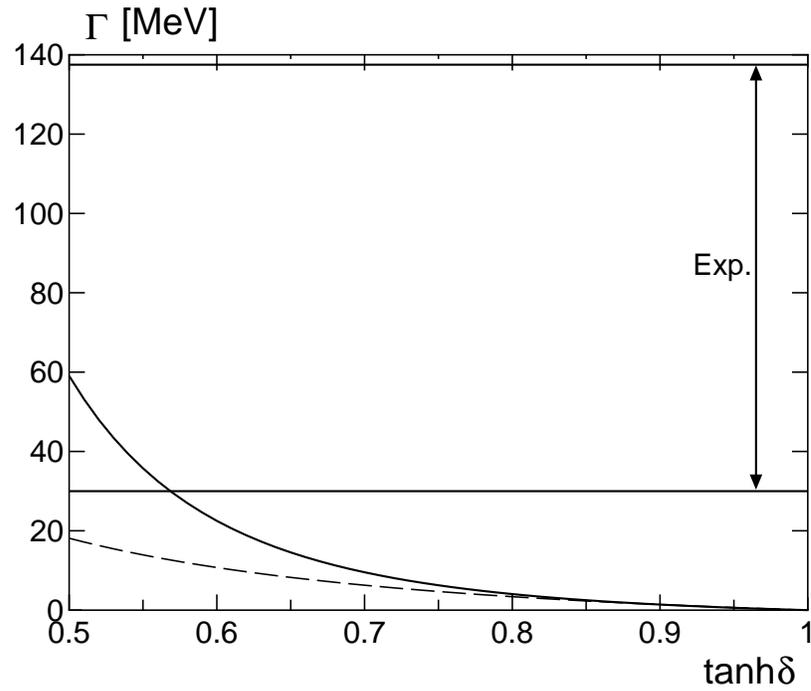}
        \caption{The same as Fig.3 for $N_- \to N \eta$}
\end{figure}

\begin{figure}[t]
	\hspace*{3cm}
	\epsfxsize=310pt
 	\epsfbox{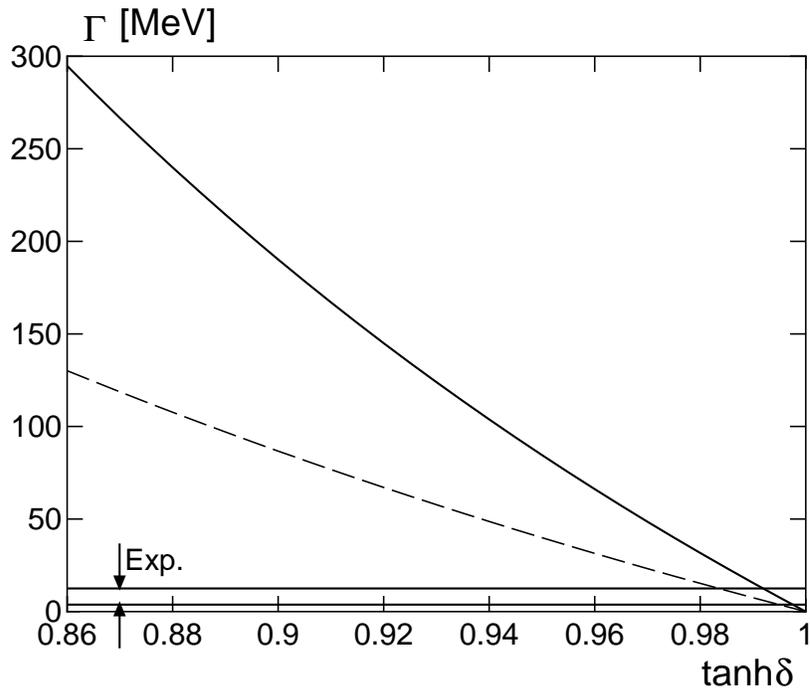}
	\caption{The same as Fig.3 for $\Lambda_- \to N \bar{K}$}
\end{figure}

\begin{figure}[h]
	\hspace*{3cm}
        \epsfxsize=310pt
        \epsfbox{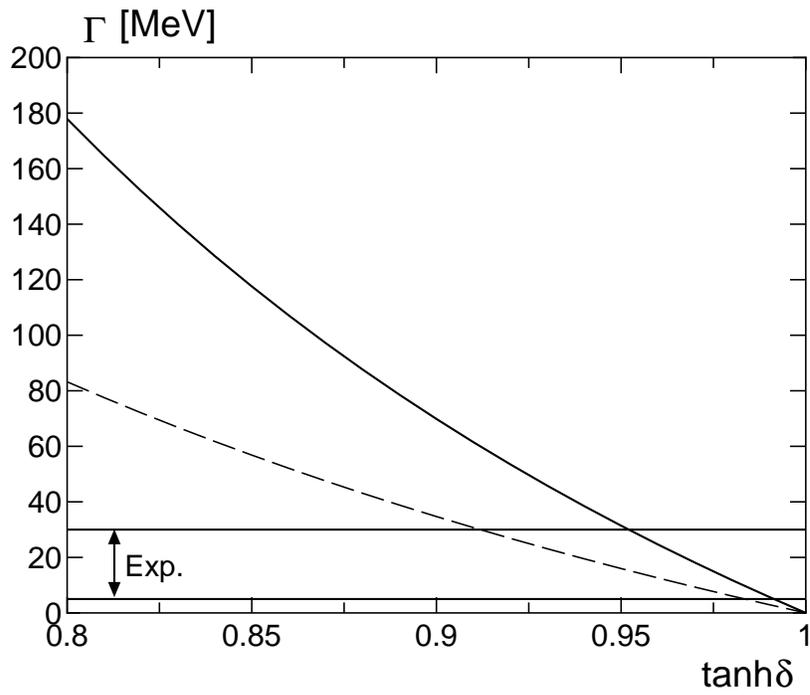}
        \caption{The same as Fig.3 for $\Lambda_- \to \Sigma \pi$}
\end{figure}

\begin{figure}[t]
	\hspace*{3cm}
        \epsfxsize=310pt
        \epsfbox{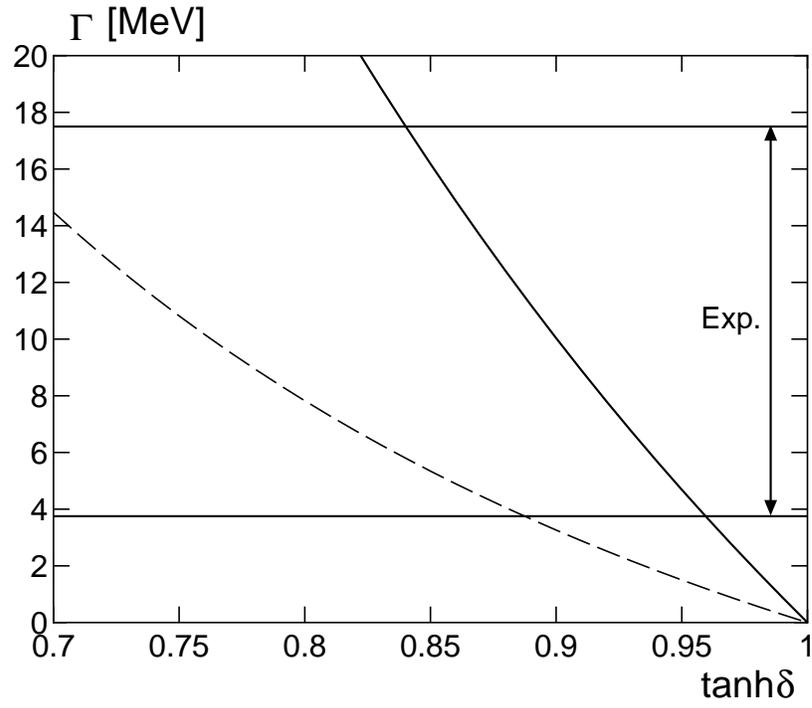}
        \caption{The same as Fig.3 for 
                 $\Lambda_- \to \Lambda \eta$}
\end{figure}

\begin{figure}[h]
	\hspace*{3cm}
        \epsfxsize=310pt
        \epsfbox{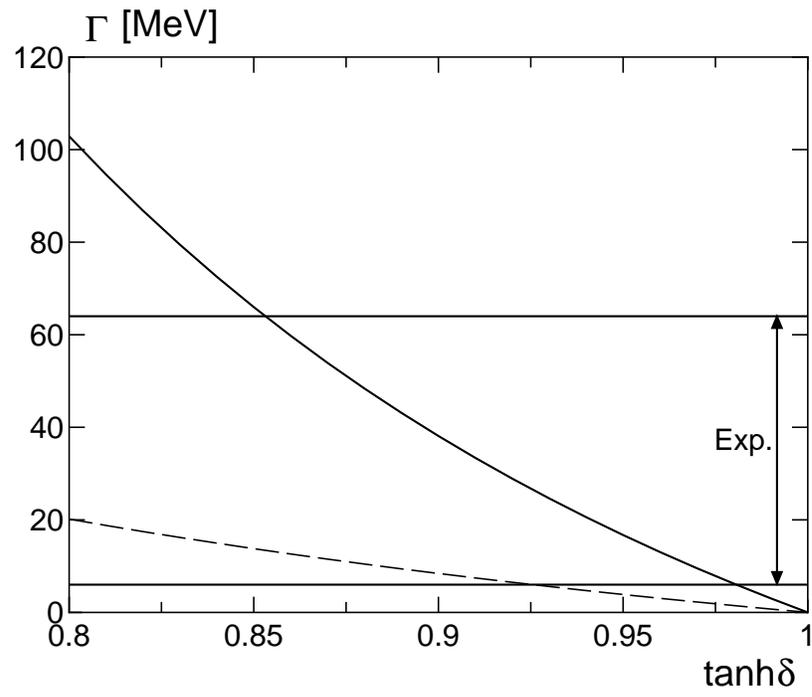}
        \caption{The same as Fig.3 for $\Sigma_- \to N \bar{K}$}
\end{figure}

\begin{figure}[t]
	\hspace*{3cm}
        \epsfxsize=310pt
        \epsfbox{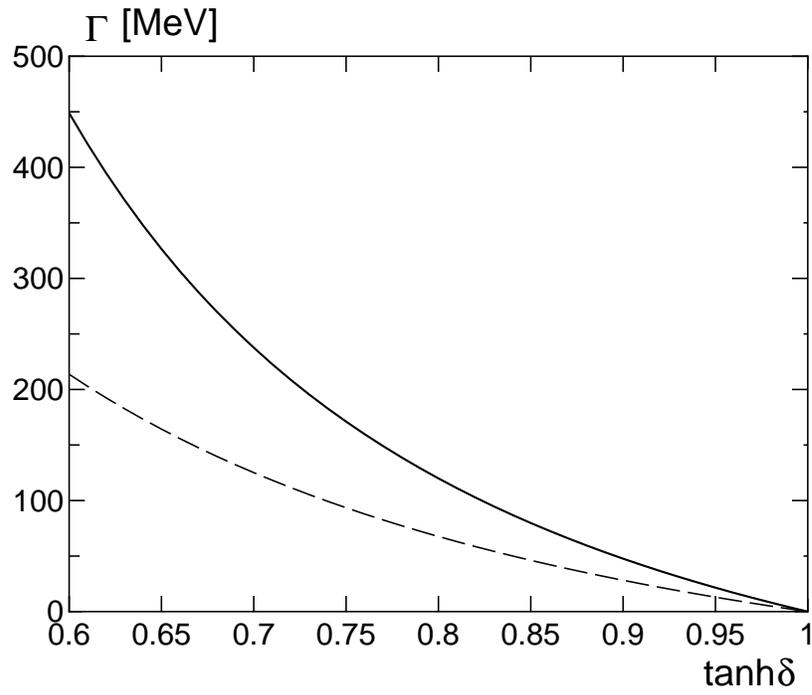}
        \caption{The same as Fig.3 for $\Sigma_- \to \Lambda \pi$}
\end{figure}

\begin{figure}[h]
	\hspace*{3cm}
        \epsfxsize=310pt
        \epsfbox{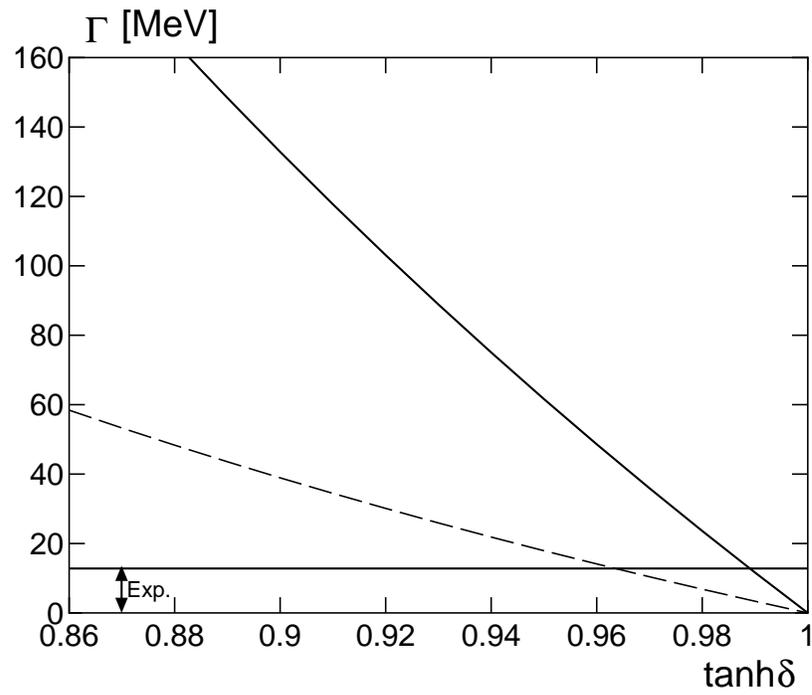}
        \caption{The same as Fig.3 for $\Sigma_- \to \Sigma \pi$}
\end{figure}

\begin{figure}[t]
	\hspace*{3cm}
        \epsfxsize=310pt
        \epsfbox{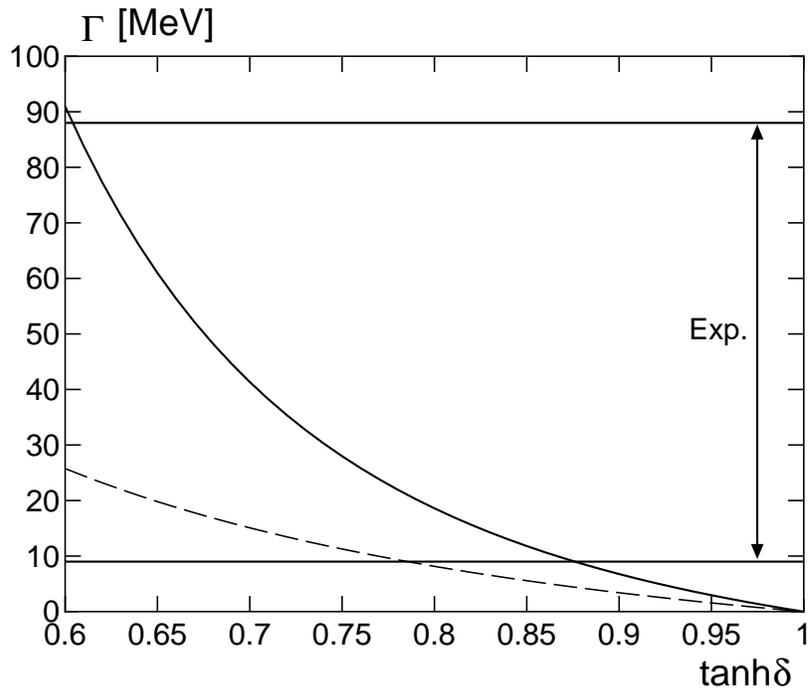}
        \caption{The same as Fig.3 for $\Sigma_- \to \Sigma \eta$}
\end{figure}

\begin{figure}[h]
	\hspace*{3cm}
        \epsfxsize=310pt
        \epsfbox{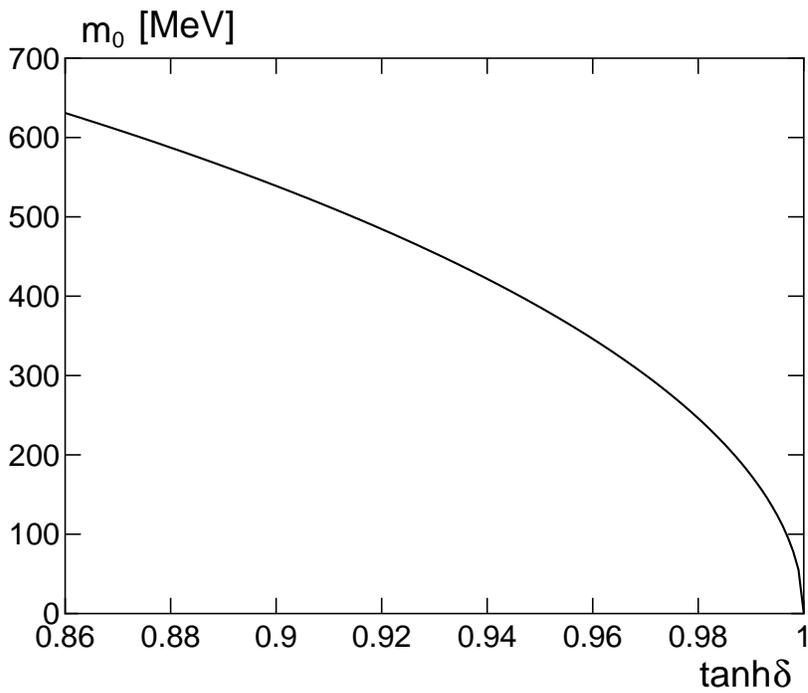}
        \caption{$\delta$ dependence of the parameter $m_0$}
\end{figure}

\begin{figure}[t]
	\hspace*{4cm}
	\epsfxsize=200pt
	\epsfbox{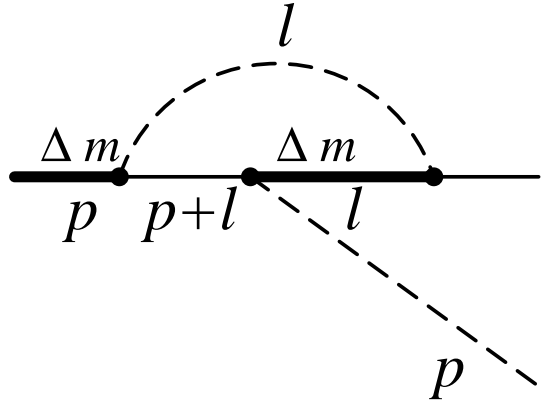}
	\caption{An example of one-loop diagram}
	\label{fig:app}
\end{figure}

\begin{figure}[h]
	\hspace*{4cm}
	\epsfxsize=200pt
	\epsfbox{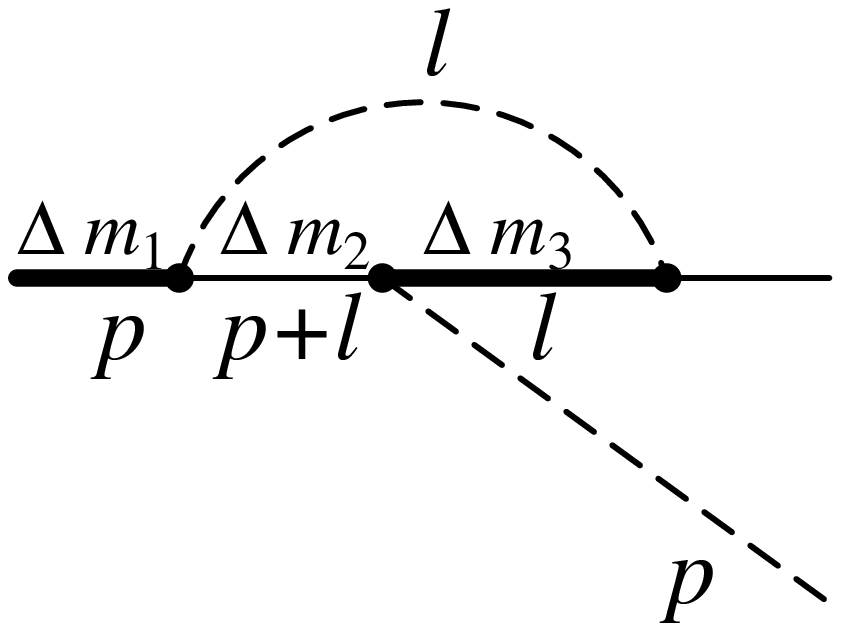}
	\caption{The same diagram as Fig.13 with the $SU(3)$
                 breaking masses of baryons}
	\label{fig:app2}
\end{figure}


\begin{thebibliography}{99}

\bibitem{JO96a}
D. Jido, N. Kodama and M. Oka, Phys. Rev. D {\bf 54}(1996)4532; \\
D. Jido and M.Oka, hep-ph/9611322; \\
M. Oka, D. Jido and A. Hosaka, in {\it Proceedings of the Fourth CEBAF/INT 
Workshop: $N^*$ Physics,} Seattle, USA, 1996, edited by T.-S.H. Lee and 
W. Roberts (World Scientific, Singapore, 1997), p. 253.  

\bibitem{DK89}
C. DeTar and T. Kunihiro, Phys. Rev. D {\bf 39}(1989)2805.

\bibitem{Ch83}
G.A. Christos, Z. Phys. {\bf C 21}(1983)83.

\bibitem{Ch85}
G.A. Christos, Z. Phys. {\bf C 29}(1985)361;
Phys. Rev. D {\bf 35}(1987)330.

\bibitem{Zh91}
H. Zheng, CERN-TH.6327/91 (1991).

\bibitem{Zh92}
H. Zheng, CERN-TH.6522/92 (1992).

\bibitem{Mo87}
I. Montvay, Phys. Lett. {\bf B 199}(1987)89.

\bibitem{JO96b}
D. Jido, M. Oka and A. Hosaka, hep-ph/9707307; \\
A. Hosaka, D. Jido and M. Oka, hep-ph/9702294; \\
D. Jido, M. Oka and A. Hosaka, Soryushiron Kenkyu(Kyoto) 95(1997)D 69.

\bibitem{Ge84}
H. Georgi, {\em Weak Interaction and Modern Particle Theory},
(Benjamin/Cummings, Menlo Park, 1984).

\bibitem{CJ97}
T.D. Cohen and X. Ji, Phys. Rev. D {\bf 55}(1997)6870.

\bibitem{JM91a}
E. Jenkins and A.V. Manohar, Phys. Lett. {\bf B 255}(1991)558.

\bibitem{MG}
A. Manohar and H. Georgi, Nucl. Phys. {\bf B 234}(1984)189.

\bibitem{PDG}
Particle Data Group, Phys. Rev. D {\bf 54}(1996)1.

\bibitem{JM91b}
E. Jenkins and A.V. Manohar, Phys. Lett. {\bf B 259}(1991)353.


\end{thebibliography}
\end{document}